**Map-based cloning of the gene *Pm21* that confers broad spectrum resistance to wheat powdery mildew**

Huagang He[1]*, Shanying Zhu[2], Yaoyong Ji[1], Zhengning Jiang[3], Renhui Zhao[3], Tongde Bie[3]*

*Corresponding authors. E-mail: hghe@mail.ujs.edu.cn; btd@wheat.org.cn

[1]School of Food and Biological Engineering, Jiangsu University, Zhenjiang, Jiangsu 212013, China;

[2]School of Environment, Jiangsu University, Zhenjiang, Jiangsu 212013, China;

[3]Yangzhou Academy of Agricultural Sciences/Key Laboratory of Wheat Biology and Genetic Improvement on Low & Middle Yangtze River Valley Wheat Region (Ministry of Agriculture), Yangzhou, Jiangsu 225007, China.

**Abstract:** Common wheat (*Triticum aestivum* L.) is one of the most important cereal crops. Wheat powdery mildew caused by *Blumeria graminis* f. sp. *tritici* (*Bgt*) is a continuing threat to wheat production. The *Pm21* gene, originating from *Dasypyrum villosum*, confers high resistance to all known *Bgt* races and has been widely applied in wheat breeding in China. In this research, we identify *Pm21* as a typical coiled-coil, nucleotide-binding site, leucine-rich repeat gene by an integrated strategy of resistance gene analog (RGA)-based cloning via comparative genomics, physical and genetic mapping, BSMV-induced gene silencing (BSMV-VIGS), large-scale mutagenesis and genetic transformation.

Common wheat (*Triticum aestivum* L.) is the most widely grown cereal crop occupying ~17% of all cultivated land worldwide and providing ~20% of the calories consumed by humankind (Fu et al. 2009). However, wheat production is challenged constantly by powdery mildew, which is caused by *Blumeria graminis* f. sp. *tritici* (*Bgt*). Utilization of powdery mildew resistance (*Pm*) genes is an effective and economical way to reduce yield losses caused by *Bgt*. Up to now, more than 100 *Pm* genes in wheat and its relatives have been documented (McIntosh et al. 2017). Among them, *Pm21* that originates from *Dasypyrum villosum* confers a high level of resistance to all known *Bgt* races (Chen et al. 1995; Cao et al. 2011). It is important to clarify the genetic basis and functional mechanism of *Pm21*. Previously, several candidate genes, including *Stpk-V* and *DvUPK* located in chromosome 6VS bin FL0.45-0.58 carrying *Pm21*, were reported to be required by *Pm21* resistance (Cao et al. 2011; He et al. 2016); however, due to lack of a fine map, the relationships of these candidate genes and *Pm21* are unclear.

In the present study, four *D. villosum* lines (DvSus-1 ~ DvSus-4) susceptible to *Bgt* isolate YZ01 at the seedling stage were identified from a total of 110 accessions (Fig. 1E and Table S1). Fine genetic mapping of *Pm21* was conducted on an $F_2$ population derived from a cross between resistant line DvRes-1 carrying *Pm21* and susceptible line DvSus-1. Among the total 10,536 $F_2$ plants, 64 recombinants between markers 6VS-00.1 and Xcfe164 (Qi et al. 2010) on 6VS were identified. *Pm21* was then mapped to a 0.01-cM interval flanked by the markers 6VS-08.4b and 6VS-10b (Fig. 1A and Fig. S1), in which, genes *DvEXO70* (6VS-08.8b; encoding a putative exocyst complex component EXO70A1-like protein) and *DvPP2C* (6VS-09b) co-segregated with *Pm21* (Fig. S2), whereas candidate genes reported previously, such as *Stpk-V* (Cao et al. 2011), were not.

A conserved coiled-coil, nucleotide-binding site, leucine-rich repeat (CC-NBS-LRR)-encoding resistance gene analog (RGA) locus was found between wheat and *Brachypodium* by comparative mapping (Fig. 1B, 1C and 1D). Subsequently, a 17,732-bp genomic sequence harboring three complete genes, viz., *DvPP2C*, encoding a protein phosphatase (He et al. 2016), *DvRGA2* and *DvRGA1*, were obtained together with a separated gene *DvRGA3* from the resistant *D. villosum* line DvRes-1 by PCR (Fig. S3). Genetic analysis demonstrated that all the above RGAs, *DvRGA1* (6VS-09.6b), *DvRGA2* (6VS-09.4b) and *DvRGA3* (6VS-09.8b) co-segregated with *Pm21* (Fig. 1A, Fig. S1 and Fig. S2). Further physical mapping, by using the susceptible deletion line Y18-S16 identified from an EMS-induced Yangmai 18 population, showed that the entire genetic interval carrying *Pm21* was missing in Y18-S16 (Fig. S1).

The genomic sequence of *DvRGA1* is 2,986 bp in length with 2 exons and 1 intron, and the corresponding open reading frame (ORF) is 2,736 bp. The nucleotide sequence of *DvRGA2* spans 3,699 bp, harboring 3 exons and 2 introns, with a 2,730-bp ORF. The exon-intron structure of *DvRGA3* was similar to that of *DvRGA2* but there was an additional repeat sequence in the second intron and several premature stop codons in the third exon (Fig. S4), suggesting that *DvRGA3* is a pseudogene. Transcriptional analysis demonstrated that *DvRGA1* and *DvRGA2* were transcribed in *D. villosum* seedlings whereas *DvRGA3* was not. Quantitative real-time RT-PCR (qPCR) showed that *DvRGA1* and *DvRGA2* were both enhanced at the transcriptional level following *Bgt* infection, and they shared similar transcription patterns during infection of Yangmai 18 (Fig. S5).

To confirm which RGA(s) corresponded to *Pm21*, the *DvRGA1* and *DvRGA2* alleles were cloned from susceptible *D. villosum* lines DvSus-1 ~ DvSus-4 as well as susceptible wheat addition line DA6V#1 (Sears 1953; Qi et al, 1998) and then sequenced. All the five susceptible lines had variations in the *DvRGA2* gene. Interestingly, DvSus-2 and DvSus-3 had a common mutation and DvSus-4 and DA6V#1 shared another mutation in *DvRGA2*. DvSus-1 ~ DvSus-3 had abnormal *DvRGA1* alleles whereas DvSus-4 and DA6V#1 did not (Table S2). The silencing effects of *DvRGA1* and *DvRGA2* in Yangmai 18 were analyzed using BSMV-VIGS technology. Silencing of *DvRGA2* allowed normal development of powdery mildew with macroscopic disease symptoms and sporulation on leaves at 10 days post-inoculation (Fig. 2A and 2B). However, no obvious effects were observed after silencing of *DvRGA1* and *DvEXO70* (Fig. S6). Silencing of *DvPP2C* led to sporulation but without significant macroscopic disease development (He et al. 2016). These results indicated that *DvRGA2* is required for *Pm21* resistance.

To detect if *DvRGA2* is sufficient for the resistance, *DvRGA2* with its native promoter was transformed into susceptible cv Kenong 199 by particle bombardment. Six of the 16 $T_1$ families identified with markers MBH1 and 6VS-09.4 were positively transgenic and showed immunity to *Bgt* isolate YZ01 and to *Bgt* isolate mixtures collected from different regions (Fig. 2C). We concluded that *DvRGA2* expressed by its native promoter confers resistance to powdery mildew in wheat.

To verify whether *DvRGA2* is *Pm21*, Yangmai 18 carrying *Pm21* was mutagenized with EMS. Fifty eight independent susceptible mutants were identified among 6,408 $M_2$ families (Fig. 2D). Except for Y18-S16, which had a deletion spanning the *Pm21* locus, each of the other 57 susceptible mutants harbored a mutated *DvRGA2* sequence. Among them, 55 each had a single-base mutation in *DvRGA2*, whereas the other two, Y18-S35 and Y18-S43, each had two-base changes. Sixteen of the 59 mutation sites caused premature stop codons and 43 caused changes in amino acids (Fig. 2E, Table S3). We also checked the flanking genes *DvPP2C* (He et al. 2016) and *DvRGA1* in 11 randomly selected mutants, and found no differences from that in untreated Yangmai 18. We concluded that *DvRGA2* alone is *Pm21*.

Amino acid sequence analysis showed that DvRGA2 protein had relatively high identities with BRADI3G03874, BRADI3G03878, BRADI3G03882 and BRADI3G03935 (51.2 ~ 60.5%). Among the CC-NBS-LRR proteins reported in wheat, DvRGA2 shared highest identity (35.0%) with stem rust resistance proteins SR22 but lower identities with powdery mildew resistance proteins PM2 (15.2%), PM3b (18.7%) and PM8 (19.8%) (Fig. S7). We searched for *Pm21* orthologs in the wheat genome and found that they are present but disrupted by a transposon-like element (6AS) or not completely assembled (6BS and 6DS) in the second intron. *Pm21* orthologs in other related species, such as *T. urartu*, *Aegilops speltoides* and *Ae. tauschii*, the donors of wheat subgenomes, are all disrupted by transposon-like elements (Fig. S8). It appears that the events causing structural abnormalities of *Pm21* orthologs occurred after divergence of *D. villosum* and the other species.

In the past, it was extremely difficult to clone *Pm21* in wheat background by map-based strategy due to lack of recombination between alien chromosome 6VS and wheat homoeologous chromosomes. The present break-through came with the discovery of several powdery mildew susceptible *D. villosum* lines, allowing construction of a high density genetic map of chromosome 6VS within that species. Although several genes had been reported to be required for *Pm21* resistance (Cao et al. 2011; He et al. 2016), and even overexpression of *Stpk-V* conferred high resistance to powdery mildew in transgenic wheat, none apart from *DvPP2C* was located in the genetic interval carrying *Pm21*.

Although fine genetic mapping allows gene isolation by using the map-based cloning, the method is usually dependent on a bacterial artificial chromosome (BAC) library that is time consuming, labour intensive and expensive to develop. Given that most of the identified disease resistance (*R*) genes in wheat, such as *Pm2* (Sánchez-Martín et al. 2016), *Pm3* (Yahiaoui et al. 2004), *Lr10* (Feuillet et al. 2003), *Sr33* (Periyannan et al. 2013) and *Sr35* (Saintenac et al. 2013), encode CC-NBS-LRR proteins, we preferentially isolated CC-NBS-LRR-encoding resistance gene analogs (RGAs) by PCR according to a conserved RGA locus common to wheat and *Brachypodium*. Among the candidates isolated,

*DvRGA2* was shown to be identical to *Pm21* via BSMV-VIGS, genetic transformation and large-scale mutagenesis.

Broad spectrum resistances are commonly controlled by non-NBS-LRR genes, such as *Yr36* (Fu et al. 2009), *Lr34* (Krattinger et al. 2009) and *Lr67* (Moore et al. 2015), rather than by NBS-LRR-encoding genes that confer race-specific resistance, such as *Pm2* (Sánchez-Martín et al. 2016) and *Pm3* (Yahiaoui et al. 2004). Nevertheless, researches show that several NBS-LRR-encoding genes can confer broad spectrum resistances, such as potato late blight resistance gene *RB* (Song et al. 2003) and rice blast resistance gene *Pi9* (Qu et al. 2006). Here, we demonstrated that the broad spectrum resistance of *Pm21* is also conferred by a single CC-NBS-LRR-encoding gene. It was proposed that *Pm21* is a relatively ancient *Pm* gene, whose product may perceive a conserved effector(s) from different *Bgt* races.

Since *Pm21* is a single CC-NBS-LRR-encoding gene, the question arises as to whether it will continue to confer durable resistance to powdery mildew. Analysis of nine independent mutations in *Pm21* revealed single amino acid changes in the LRR domain that were correlated with loss of resistance to *Bgt* isolate YZ01. Among them, the mutations in Y18-S9 and Y18-S20 (Table S3) involved changes in solvent-exposed LRR residues that are considered to control specific recognition of the pathogen (Meyers et al. 1998; Wulff et al. 2009). In the recent years, wheat varieties carrying *Pm21* are increasingly being planted in China (Bie et al. 2015), which would accelerate *Bgt* evolution, and the risk of losing *Pm21* resistance would arise. So, it will be a great challenge to maintain the resistance of *Pm21* in the future. One practical way may be pyramiding other *Pm* gene(s) into wheat varieties carrying *Pm21*.

**Materials and Methods**

**Plant materials and pathogen inoculation**

A total of 110 accessions of *Dasypyrum villosum* were collected from the Germplasm Resources Information Network (GRIN) (51), GRIN Czech (16), Genebank Information System of IPK Gatersleben (GBIS-IPK) (35), Nordic Genetic Resource Center (NordGen) (7) and Cytogenetics Institute, Nanjing Agricultural University (CI-NAU) (1). The susceptible addition line DA6V#1 (Sears 1953; Qi et al. 1998) was provided by GRIN (Table S1). Powdery mildew resistant wheat cultivar (cv) Yangmai 18 carrying a pair of translocated T6AL.6VS chromosomes (*Pm21*) and susceptible cv

Yangmai 9 were developed at Yangzhou Academy of Agricultural Sciences (YAAS). All plants were inoculated with *Bgt* isolate YZ01, a predominant isolate collected from Yangzhou (He et al. 2016), by dusting from sporulating susceptible plants and powdery mildew responses were assessed at 8 days post-inoculation. *Bgt* isolate YZ01 was maintained on cv Yangmai 9 seedlings.

**DNA isolation and development of molecular markers**

Genomic DNA was extracted from fresh leaves of seedlings by the CTAB method (Murray and Thompson 1980). DNA markers were reported previously (He et al. 2016; Qi et al. 2010; Cao et al. 2006) or newly developed using CISP and CISP-IS strategies based on collinearity among *Brachypodium*, rice and *Triticeae* species, as described by He et al. (2013). All primers used in this study are listed in Table S4.

**Genetic mapping**

An $F_2$ population was derived from the cross between the resistant *D. villosum* line DvRes-1 carrying the *Pm21* gene and seedling-susceptible line DvSus-1 (Table S1) newly found in this study. Powdery mildew responses of $F_2$ plants were determined at the one-leaf stage. For molecular analysis, PCR amplifications were performed in a Peltier Thermal Cycler (Bio-Rad, USA) in 25 μl volumes containing 1×PCR buffer, 0.2 mM of each dNTP, 2 μM of each primer, 1 unit of *Taq* DNA polymerase, and 1 μl of DNA template. PCR was carried out with an initial denaturation at 94℃ for 3 min, 35 cycles of 20 s at 94℃, 30 s at 60℃, 1 min at 72℃, and a final extension for 5 min at 72℃. PCR products were separated in 6 ~ 12% non-denaturing polyacrylamide gels, silver stained, and photographed. Chi-squared ($\chi^2$) tests were used to determine the goodness-of-fit of the observed segregation ratios to theoretical Mendelian ratios.

**PCR amplification of candidate RGAs**

Degenerative primers used for cloning of candidate genes were designed according to the conserved sequences of predicted RGAs in wheat and *Brachypodium* in the orthologous regions of the *Pm21* locus. Fragments of candidate RGAs were PCR-amplified from genomic DNA of *D. villosum* line DvRes-1 carrying *Pm21*. Thermal-asymmetric-interlaced (TAIL) PCR (Liu and Huang 1998) and Long-range (LR) PCR (Song et al. 2003) with LA *Taq* DNA polymerase (TaKaRa, Japan) were further used to clone unknown DNA fragments close to the candidate genes.

**Quantitative real-time RT-PCR (qPCR)**

Total RNA was isolated from wheat and *D. villosum* leaves inoculated or non-inoculated with *Bgt*

isolate ZY01, using TRIzol reagent (Life Technologies, USA). First-strand cDNA was then synthesized from 2 μg of total RNA using a PrimeScript™ II 1st Strand cDNA Synthesis Kit (TaKaRa, Japan). Quantitative real time RT-PCR (qPCR) was performed in an ABI 7300 Real Time PCR System (Life Technologies, USA) as described by He et al. (2016). The wheat actin gene (*TaACT*) was used as reference gene as reported (Bahrini et al. 2011). All reactions were run in three technical replicates for each cDNA sample.

**Sequence analysis**

The genome sequences of *Brachypodium*, rice and wheat were obtained from *Brachypodium distachyon* genome assemblies v2.0 (http://www.brachypodium.org), rice genome pseudomolecule release 7 (http://rice.plantbiology.msu.edu), and the IWGSC Sequence Repository (http://wheat-urgi.versailles.inra.fr), respectively. Genes were predicted using the FGENESH tool (Solovyev et al. 2006), and then re-annotated by using the BLAST program (Johnson et al. 2008) in combination with the SMART program (Letunic et al. 2015). Protein domain prediction and multiple sequence alignment analysis were performed by the SMART and CLUSTAL W (Thompson et al. 1994) tools, respectively. Phylogenetic tree was constructed by the Neighbor-Joining method in the MEGA4 software (Tamura et al. 2007).

**Functional analysis of candidate genes by BSMV-VIGS**

BSMV-VIGS (Hein et al. 2005; Scofield et al. 2005) was utilized to investigate the potential involvement of the candidate genes in wheat cv Yangmai 18. Gene fragments were amplified from the first-strand cDNA of *D. villosum*, digested with *Eco*RⅠ/*Sal*Ⅰ, and then inserted in reverse orientation into a modified BSMV: γ vector. The details of silencing of target genes were described in our previous work (He et al. 2016).

**Wheat transformation**

The vectors pAHC25 (Christensen and Quail 1996) was digested with *Hind*Ⅲ, and then the large fragment containing the *bar* gene was ligated with the multiple cloning sites (*Sma*Ⅰ- *SnaB*Ⅰ- *Eco*RⅤ- *Stu*Ⅰ- *Not*Ⅰ), generating pAHC25-MCS2. The 5,890-bp genomic DNA of *DvRGA2*, containing a 1,779-bp native promoter sequence and a 412-bp downstream sequence, was PCR-amplified using PrimeSTAR Max Premix (TaKaRa, Japan) according to the manufacturer's guideline. After digestion with *Sma*Ⅰ and *Not*Ⅰ, *DvRGA2* was inserted into pAHC25-MCS2. After confirmed by sequencing, the construct was transformed into susceptible wheat cv Kenong 199 using the PDS-1000/He biolistic particle delivery system (Bio-Rad). $T_1$ plants were tested for presence of the transgene by PCR-amplification using markers MBH1 (Bie et al. 2015) and 6VS-09.4, located in the promoter region and coding region of *DvRGA2*, respectively. Marker 6VS-09.6 derived from *DvRGA1* was also used as a negative control. The positively identified $T_1$ plants were inoculated with *Bgt* isolate YZ01 and mixed isolates collected from different regions of China.

**Mutation analysis**

About 10,000 seeds of Yangmai 18 were treated with 0.8% ethyl methanesulfonate (EMS), and 6,408 $M_2$ families were obtained. About 100 seeds of each $M_2$ family were screened for mutants that were susceptible to powdery mildew. Susceptibility of mutants was confirmed in adult plant tests. The *DvRGA2* gene in each mutant line was obtained by RT-PCR, inserted into pAHC25-MCS1 after digestion with *Sma*Ⅰ/*Spe*Ⅰ, and sequenced by the Sanger method. pAHC25-MCS1 was derived from the vector pAHC25, in which, the *gus* gene was replaced by multiple cloning sites (*Sma*Ⅰ - *Not*Ⅰ - *Mlu*Ⅰ - *Spe*Ⅰ - *Sac*Ⅰ). Each mutation was verified by sequencing the PCR product harboring the candidate mutation site. As controls, the *DvRGA1* and *DvPP2C* genes in 11 randomly selected mutants were also obtained and sequenced by the same method. The distribution of all *DvRGA2* mutations in sequences from susceptible wheat lines was analyzed at the protein domain and motif levels according to previous descriptions (Meyers et al. 1999; Dilbirligi and Gill 2003). The *DvRGA1* and *DvRGA2* genes in the susceptible and resistant *D. villosum* lines were amplified by LR PCR from genomic templates, cloned into the pMD18-T vector, and sequenced.

**Acknowledgments:** This research was supported by Grants from the National Natural Science Foundation of China (31471497), the Natural Science Foundation of Jiangsu Province (BK20130503) and the Innovation Foundation of Jiangsu Academy of Agricultural Sciences [ZX(17)2011]. The authors are grateful to the Germplasm Resources Information Network (GRIN), GRIN Czech, Genebank Information System (GBIS) of the IPK Gatersleben, Nordic Genetic Resource Center (NordGen), and Cytogenetics Institute, Nanjing Agricultural University (CI-NAU) for providing *D. villosum* accessions. The *Pm21* sequence was deposited in GenBank under accession number MF370199.

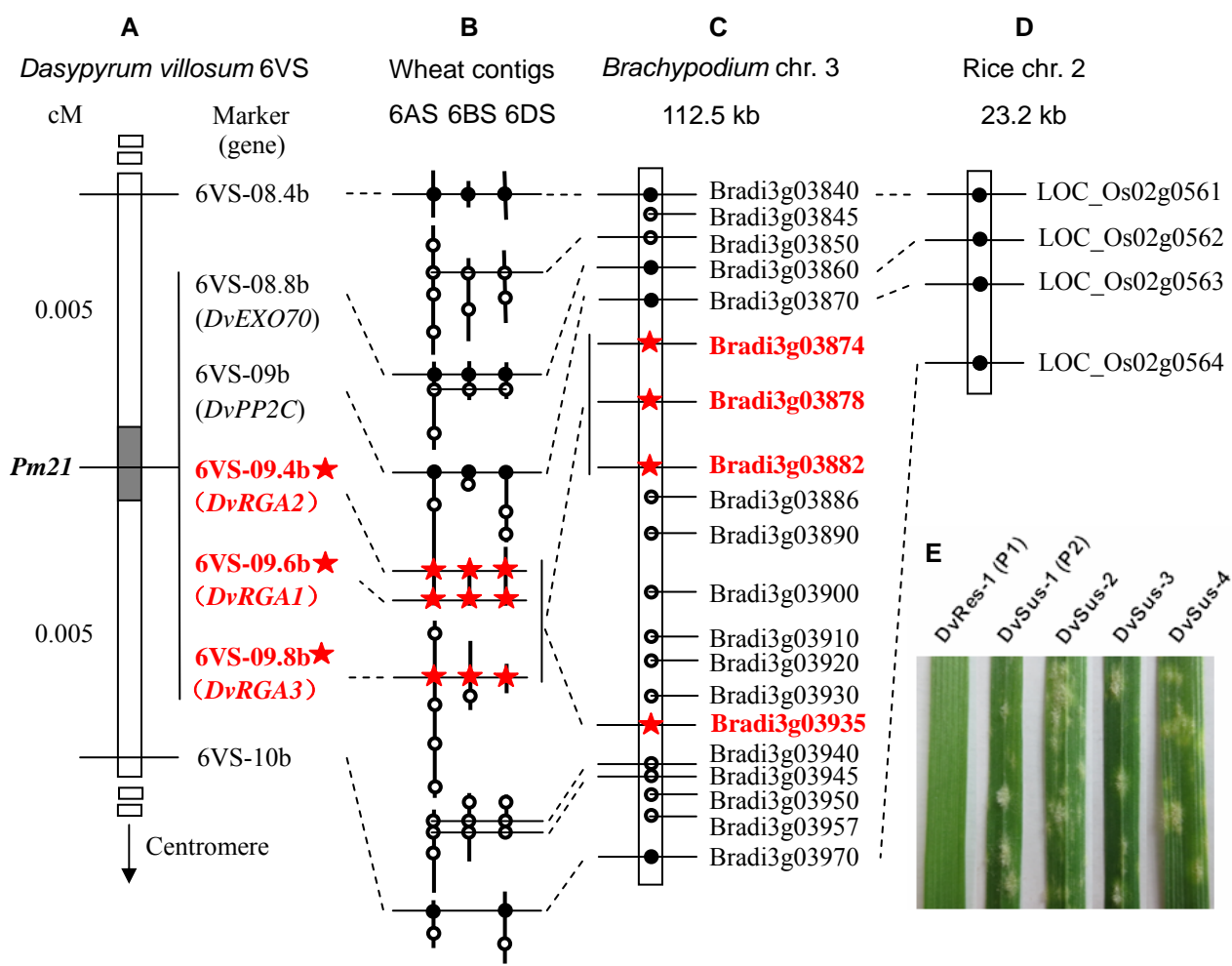

**Fig. 1. Genetic and comparative mapping of *Pm21*.** (**A**) Partial genetic map produced using an F$_2$ population derived from a cross between resistant *D. villosum* line DvRes-1 and susceptible line DvSus-1. The dark region on chromosome 6VS shows the genetic interval carrying *Pm21*. The vertical *arrow* shows the direction of the 6V centromere. The complete genetic map is showed in Fig. S1. (**B**, **C** and **D**) Comparative maps of the interval carrying *Pm21* among *D. villosum* 6VS, wheat 6AS, 6BS and 6DS (**B**), and the short arms of *Brachypodium* chromosome 3 (**C**) and rice chromosome 2 (**D**). Homologous wheat contigs of genes of interest were obtained and annotated using the BLAST, FGENESH and SMART programs. All genes from *Brachypodium* and rice are adopted according to the annotations of the corresponding genomes except RGA Bradi3g03935 in *Brachypodium* that was re-annotated in this study. Solid and hollow circles indicate conserved genes and non-conserved genes among different genomes, respectively. Orthologous RGAs among *D. villosum*, wheat and *Brachypodium* are marked by red stars. (**E**) Phenotypes of the resistant parent DvRes-1 (P1) and the susceptible parent DvSus-1 (P2) used for genetic mapping, as well as susceptible *D. villosum* lines DvSus-2 ~ DvSus-4. The powdery mildew responses were assessed following inoculation with *Bgt* isolate YZ01 at the one-leaf stage.

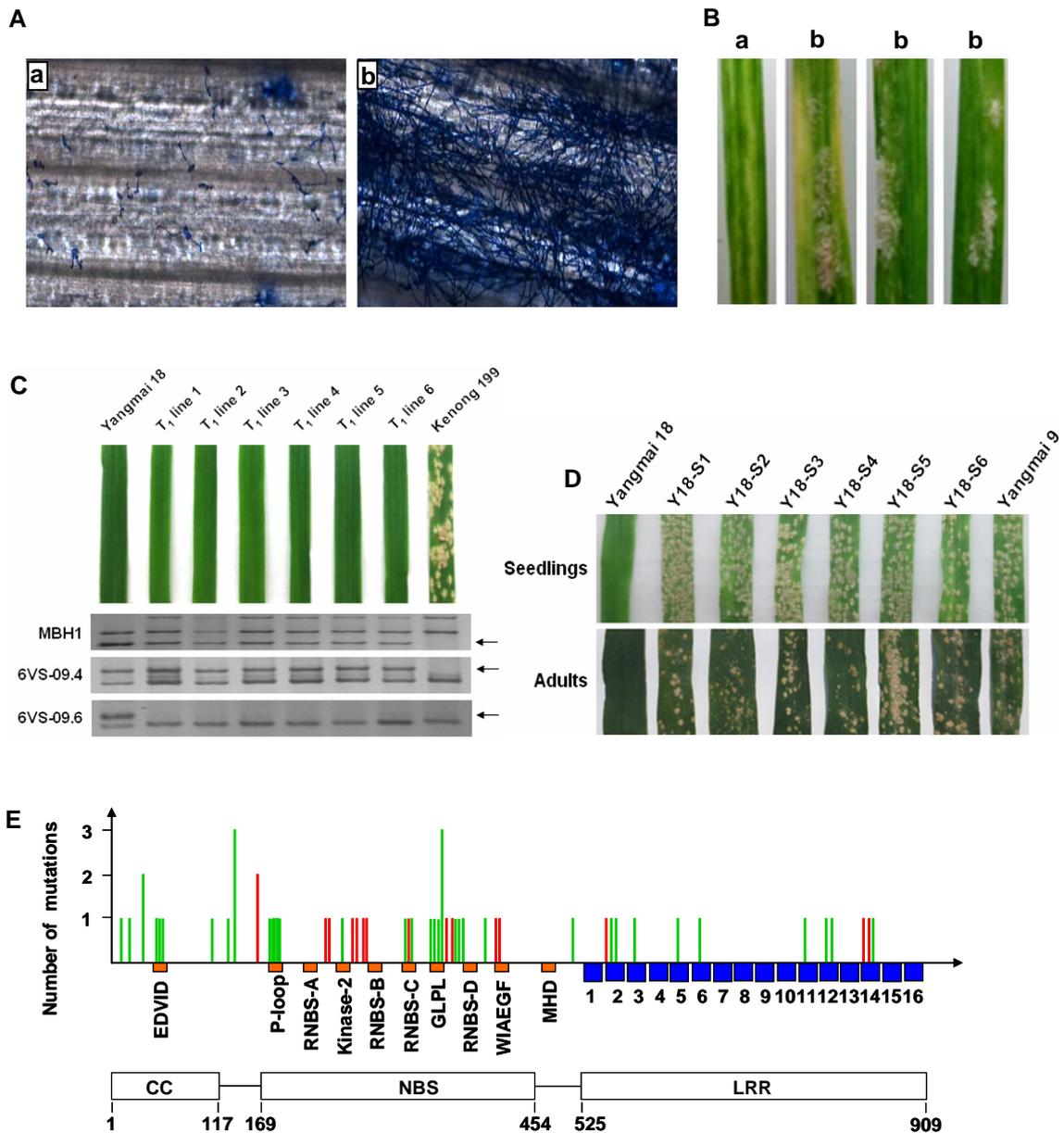

**Fig. 2. Functional validation of the *DvRGA2* gene**. (**A** and **B**) Microscopic (A) and macroscopic (B) phenotypes of the resistant wheat cv Yangmai 18 carrying *Pm21* after treatment with BSMV:00 (a) and BSMV:*DvRGA2*as (b). Fungal structures are enlarged 100 times. (**C**) Phenotypes of T$_1$ transgenic wheat lines at the seedling stage. Yangmai 18 and Kenong 199 were used as the resistant and susceptible controls, respectively. Molecular identifications of transgenes are shown below the corresponding leaves. (**D**) Phenotypes of Yangmai 18 and susceptible wheat mutants (Y18-S1 ~ Y18-S6) derived from EMS-treated Yangmai 18 at the seedling and the adult stages. Yangmai 9 was used as a susceptible control. (**E**) Frequency and distribution of mutations in *DvRGA2*. The CC, NBS and LRR domains of the putative DvRGA2 protein are showed below. The known conserved motifs

(brown) in the CC and NBS domains and 16 leucine-rich repeat motifs (blue) in the LRR domain are also shown. Mutations involved in amino acid changes and premature stop codons are shown in green and red, respectively. Variant *DvRGA2* genes in Y18-S35 and Y18-S43 with two base changes are not included because it was not clear if each change affected the reaction. Details of all mutation sites are listed in Table S3.

# Supplementary Materials

**Fig. S1 to S8**

**Tables S1 to S4**

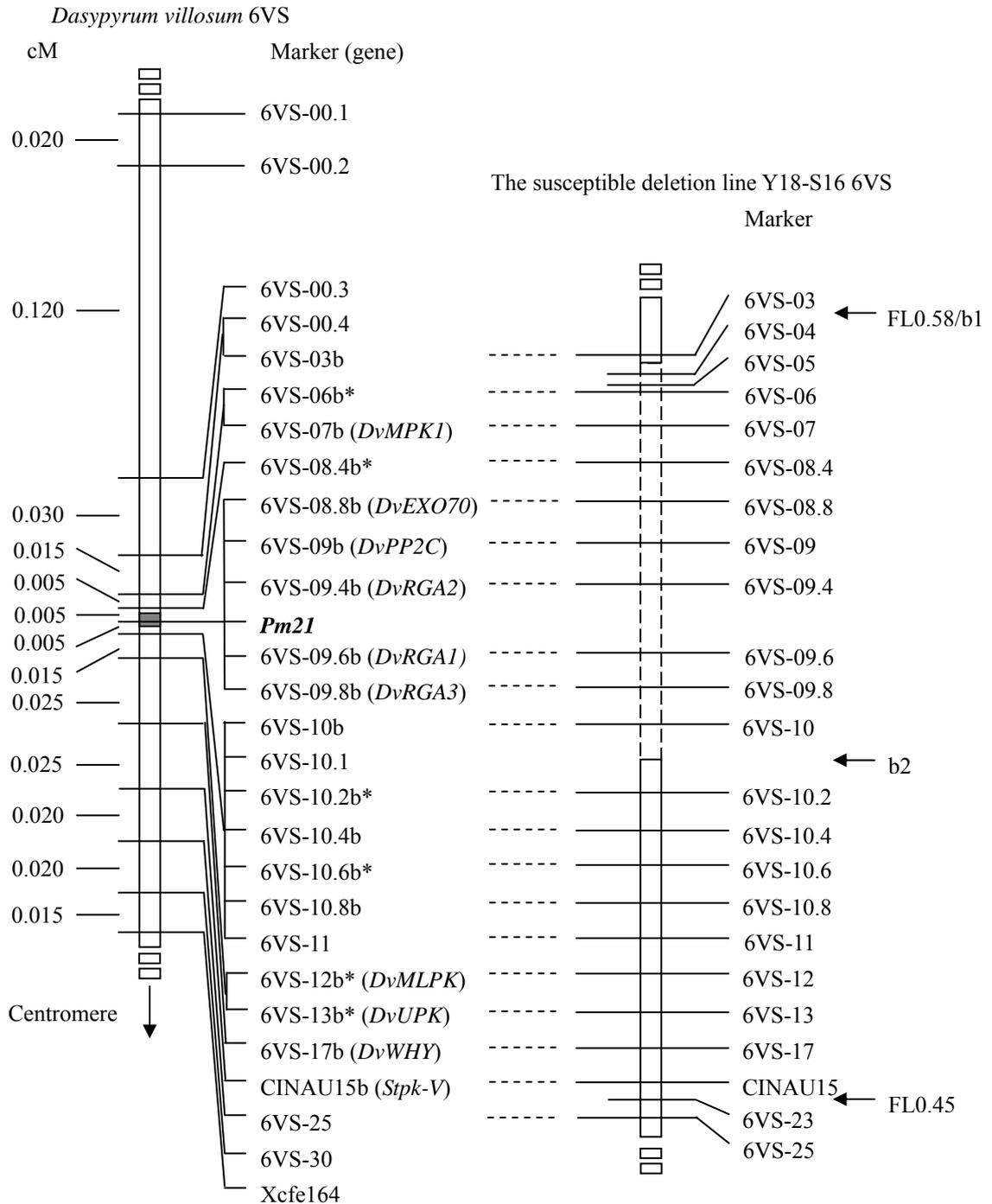

**Fig. S1**

Genetic and physical mapping of the *Pm21* gene. The genetic map of *Pm21* and 6VS markers was obtained by using an F$_2$ population derived from the cross between the resistant *D. villosum* DvRes-1 (P1) and susceptible DvSus-1 (P2) (Fig. 1E). The gray region on the chromosome 6VS marks the genetic interval carrying *Pm21*. All markers used here were derived from genes based on comparative genomics among wheat and *Brachypodium* and designated according to the gene order in *Brachypodium*. DNA markers marked by b or not were developed from the same gene. Asterisks indicate single nucleotide polymorphism (SNP) markers. The genes in brackets were reported previously (Cao et al. 2011; He et al. 2016)[4,5] or were first named in this study. The physical map was obtained using the susceptible deletion line Y18-S16. The 6VS chromosome breakpoints b1 and b2 in Y18-S16 as well as bin FL0.45-0.58 are indicated by horizontal *arrows*. Both the breakpoints b1 and FL0.58 are flanked by the markers 6VS-03 and 6VS-04. The dashed segment represents the deleted region in Y18-S16. Vertical *arrow* shows the direction of the 6VS centromere.

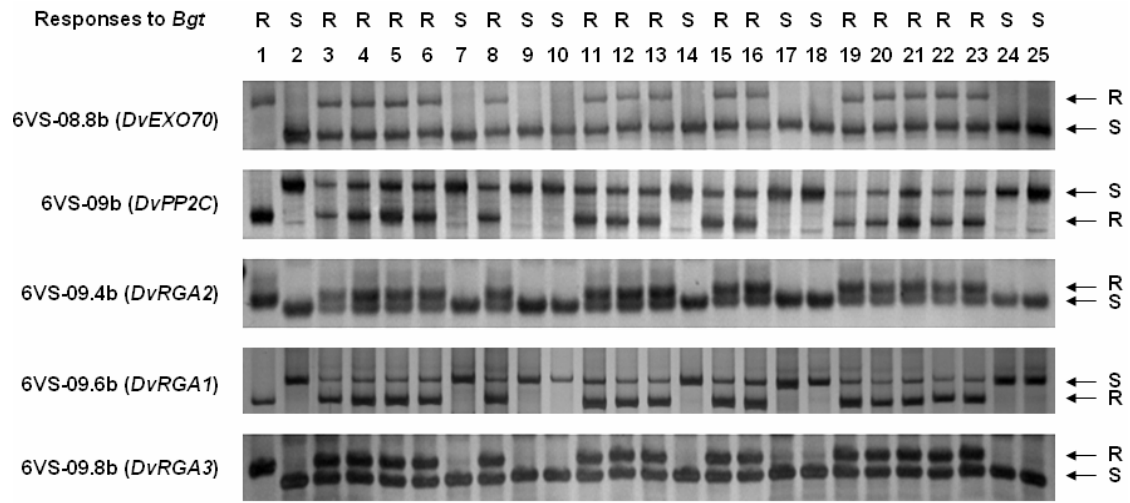

**Fig. S2**

PCR amplification patterns of co-segregating markers 6VS-08.8, 6VS-09b, 6VS-09.4 and 6VS-09.8, corresponding to the genes *DvEXO70*, *DvPP2C*, *DvRGA2* and *DvRGA3*, respectively. 1: resistant parent DvRes-1 (P1); 2: susceptible parent DvSus-1 (P2); 3 to 25: recombinants screened from the $F_2$ popution. Powdery mildew responses of the parents and recombinants are showed at the top. The *arrows* indicate DNA bands produced by resistant (R) or susceptible (S) parents.

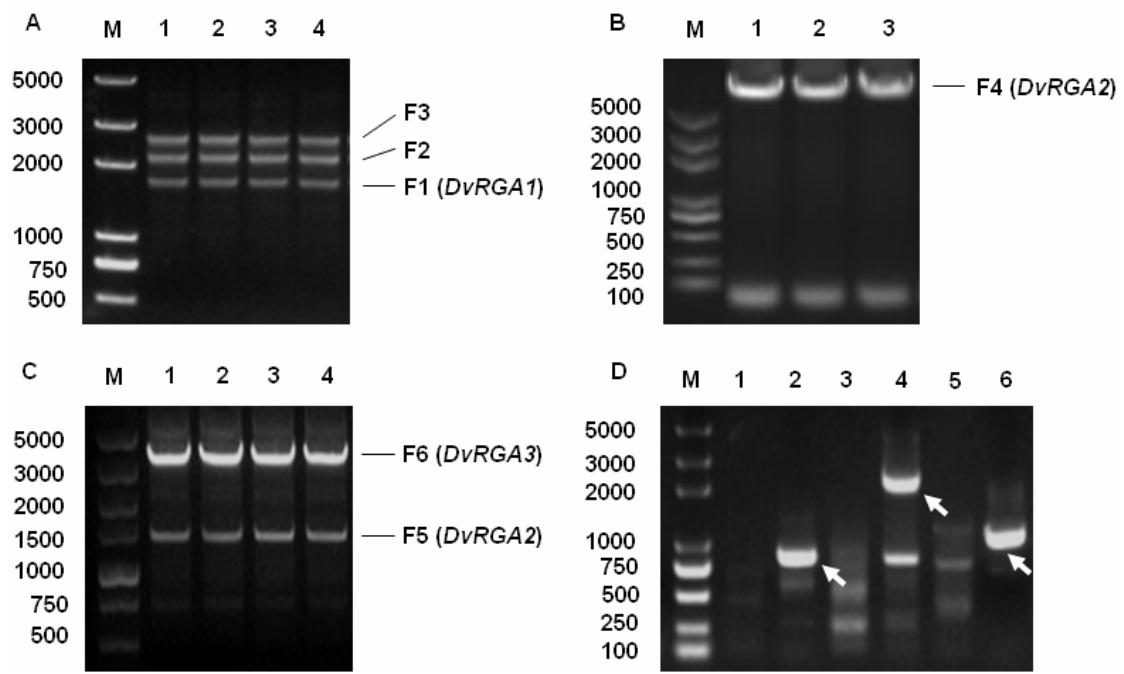

**Fig. S3**

Isolation of three candidate RGAs from *D. villosum* by PCR. (A) PCR amplification of the DNA fragment of *DvRGA1* using the degenerative primers. (B) Full-length *DvRGA2* obtained by LR PCR. (C) PCR amplification of DNA fragment of *DvRGA3*. (D) Amplification of unknown fragments near to *DvRGA1* by TAIL PCR. Specific products are indicated by *arrows*.

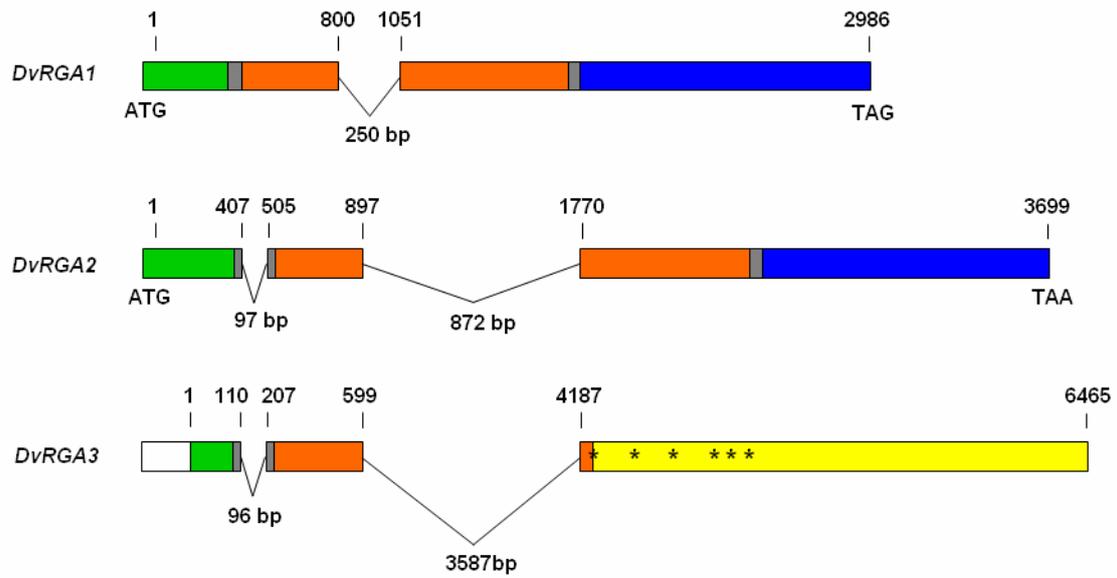

**Fig. S4**

Gene structures of *DvRGA1*, *DvRGA2* and *DvRGA3*. The green, brown and blue regions encode coiled-coil (CC), nucleotide-binding site (NBS) and leucine-rich repeat (LRR) domains, respectively. The gray regions indicate linkers between different domains. In *DvRGA3*, the white region is a putative sequence corresponding to the 5'-terminus of *DvRGA2* that could not be cloned following several attempts. Several premature stop codons in the yellow region of *DvRGA3* are marked by asterisks.

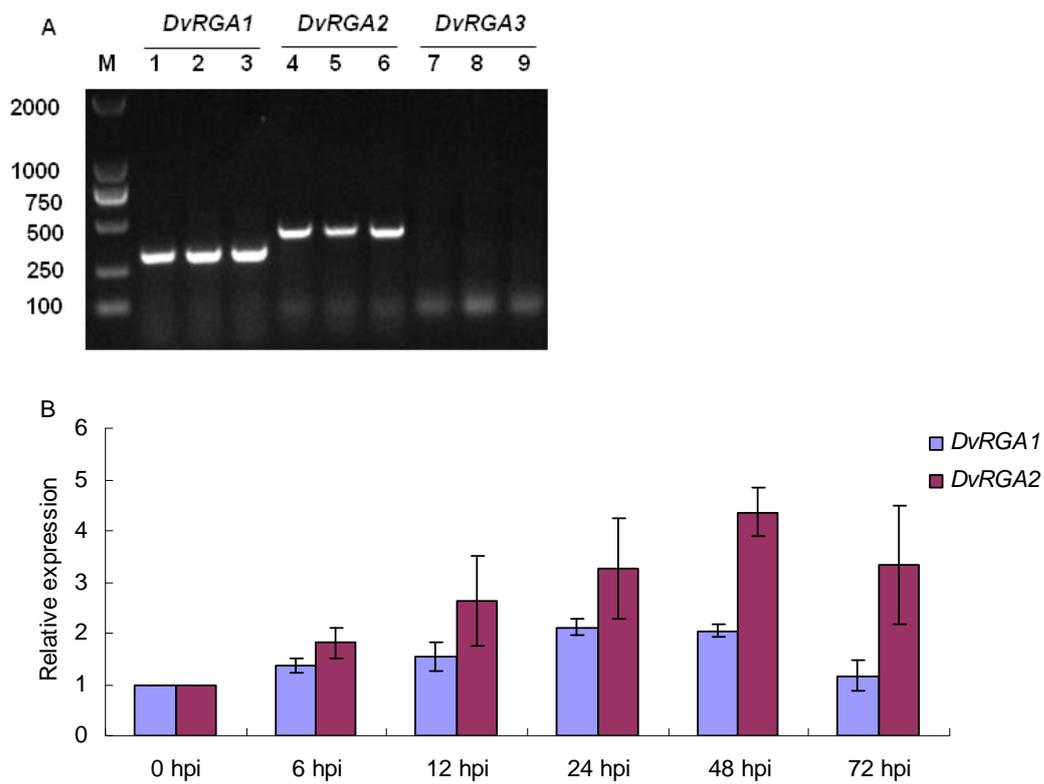

**Fig. S5**

Transcriptional analyses of candidate genes *DvRGA1*, *DvRGA2* and *DvRGA3*. (A) RT-PCR of three candidate RGAs in non-inoculated the leaves of *D. villosum*. Lanes 1 ~ 3, 4 ~ 6 and 7 ~ 9 in agrose gels are RT-PCR results for *DvRGA1*, *DvRGA2* and *DvRGA3*, respectively. (B) Quantitative real-time RT-PCR (qPCR) analysis of *DvRGA1* and *DvRGA2* in leaves of resistant wheat cv Yangmai 18 at different times post-inoculation with *Bgt* isolate YZ01.

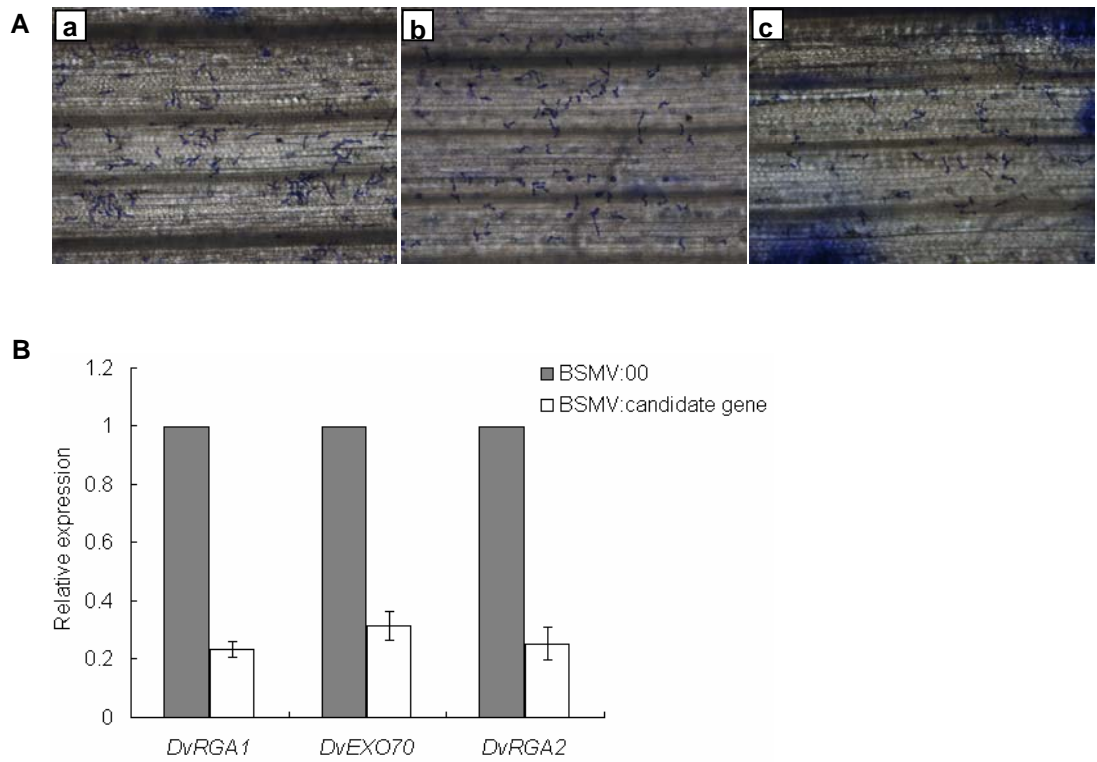

**Fig. S6**

Silencing analyses of *DvRGA1*, *DvEXO70* and *DvRGA2* co-segregating with *Pm21*. (**A**) Photomoctographs of the resistant wheat cv Yangmai 18 carrying *Pm21* after treatment with BSMV:00 (a), BSMV:*DvRGA1*as (b) and BSMV:*DvEXO70*as (c). Fungal structures are enlarged 100 times. Contrasting photomicrograph and photomoctograph of silenced *DvRGA2* are presented in Fig. 2A and 2B, respectively. (**B**) Silencing efficiencies of *DvRGA1*, *DvEXO70* and *DvRGA2* measured by qPCR, using a BSMV:00-infected sample as control.

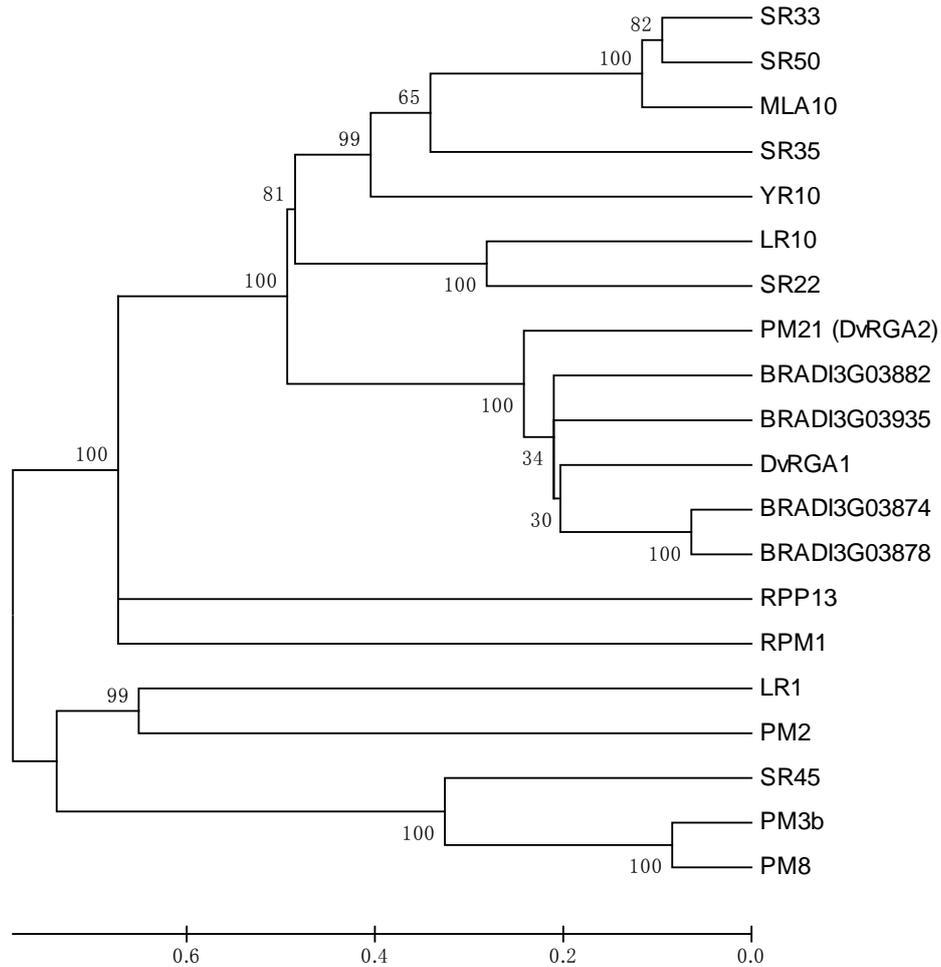

**Fig. S7**

Neighbor-joining phylogenetic tree of PM21 and relevant CC-NBS-LRR proteins, including DvRGA1, wheat powdery mildew resistance proteins PM2 (CZT14023.1), PM3b (AAQ96158.1) and PM8 (AGY30894.1), barley powdery mildew resistance protein MLA10 (AAQ55541.1), wheat stem rust resistance proteins SR22 (CUM44200.1), SR33 (AGQ17382.1), SR35 (AGP75918.1), SR45 (CUM44213.1) and SR50 (ALO61074.1), wheat leaf rust resistance proteins LR1 (ABS29034.1) and LR10 (AAQ01784.1), wheat yellow rust resistance protein YR10 (AAG42168.1), putative *Brachypodium* resistance proteins BRADI3G03874, BRADI3G03878, BRADI3G03882 and BRADI3G03935, and *Arabidopsis thaliana* resistance proteins RPM1 (NP_187360.1) and RPP13 (AAF42831.1).

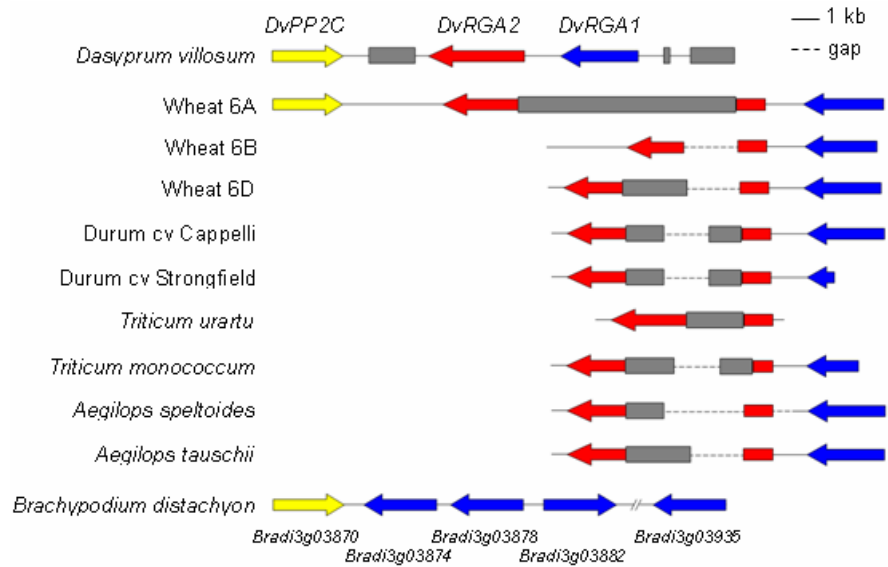

**Fig. S8**

Genomic organizations of *DvPP2C-DvRGA2-DvRGA1* at the *Pm21* locus in *D. villosum* and *DvRGA2-DvRGA1* orthologs in common wheat, wheat relatives and *B. distachyon*. All sequences from wheat species were obtained from the IWGSC Sequence Repository (http://wheat-urgi. versailles.inra.fr). They were: contig_4363243 from wheat 6AS, contig_3017519 and contig_2960485 from wheat 6BS, contig_2091261 and contig_209909 from wheat 6BS, contig_183955 and contig_285456 from *T. durum* cv Cappelli, contig_502242 and contig_2184681 from *T. durum* cv Strongfield, contig_341018 from *T. urartu*, contig_921902 and contig_2216578 from *T. monococcum*, contig_1596045, contig_290492 and contig_1586106 from *Aegilops speltoides*, and contig_98669 and contig_155986 from *A. tauschii*. The *Brachypodium* genes *Bradi3g03870*, *Bradi3g03874*, *Bradi3g03878* and *Bradi3g03882* are annotated in the *Brachypodium* genome assemblies v2.0 (http://www.brachypodium.org), whereas *Bradi3g03935* was re-annotated in this study. The red and blue *arrows* indicate *DvRGA2*-like and *DvRGA1*-like genes, respectively. The gray boxes represent transposons or other repeat sequence in wheat species. The dashed lines are unknown gaps.

**Table S1** Details of *D. villosum* lines and the wheat addition line DA6V#1 used in this study.

**Table S2** Detection of the *DvRGA1* and *DvRGA2* alleles in susceptible *D. villosum* lines and susceptible addition line DA6V#1.

**Table S3** Mutations in the *DvRGA2* gene in susceptible Yangmai 18 mutants.

**Table S4** Primers and DNA markers used in this study.